\documentstyle[12pt,a4]{article}
\begin{document}
\title{Intensity Fluctuations in Closed and Open Systems}
\author{ R. Pnini and B. Shapiro \\  \\                         
\small Department of Physics \\
\small Technion --- Israel Institute of Technology \\
\small 32000 Haifa, Israel}
\date{\small \today}      
\maketitle
\begin{abstract}
\noindent We consider the intensity pattern, generated by a
monochromatic
source, in a disordered cavity coupled to the environment. For 
weak coupling, and when the source frequency is  tuned to a 
resonance, the intensity distribution $P(I)$ is close to
Porter-Thomas distribution. When the coupling increases, $P(I)$ 
gradually crosses over to the Rayleigh distribution. The joint
probability distribution for intensities at two different points
is  also discussed.
\end{abstract}
PACS numbers: 42.20-y, 71.55Jv
\baselineskip 28pt plus 2pt

A wave propagating in a random medium
produces a complicated, highly irregular intensity pattern.
This pattern is described in statistical terms. One considers  
an ensemble of different realizations of the random medium 
and inquires, e.g., about the probability distribution $P(I)$
of the wave intensity $I$ at some point ${\bf{r}}$. In many 
cases $P(I)$ is accurately  described by the Rayleigh
distribution
\begin{equation}
P_{\scriptscriptstyle \!\! R}(I)=
\frac{1}{\langle{I}\rangle}\exp(-I/\langle{I}\rangle) \ \ ,
\end{equation}ý
where $\langle{I}\rangle$ is the average intensity.
A simple derivation of Eq.~(1) is based on an
assumption that the field at point ${\bf{r}}$ can be viewed
as a sum of many random contributions\cite{ish,ebe}. 
A more systematic derivation, which
paves the way for calculating corrections to the Rayleigh
distribution\cite{bor,cor}, is based on the perturbative
diagrammatic technique.
Here one starts with the wave equation
\begin{equation}
\left[ \nabla^2 + 
k_o^2\left(1 + \mu ({\bf{r}}) \right)
\right]
\psi_{\omega}({\bf{r}}) = 0 \ \ ,
\end{equation}
supplemented by appropriate sources. In this equation 
$\psi_\omega ({\bf{r}})$ describes a field (for instance,
pressure field in an acoustic wave or a component of 
the electric field in an electromagnetic wave), excited
by a monochromatic source of frequency $\omega$. The random
function $\mu ({\bf{r}})$ describes the fluctuating part
of the refraction index and  $k_o = \omega/ c$\ , $c$ 
being the the speed of
propagation in the average medium. 
In the diagrammatic approach one computes moments
of the intensity $I\equiv |\psi_\omega ({\bf{r}}) |^2$
and reconstructs the distribution $P(I)$.

Let us stress that Eq.~(1) applies to the case of a
monochromatic wave propagating
in an open system. A different type of problem arises if one 
considers the wave 
equation~(2)
in a closed geometry without sources.
In this case one inquires about the statistical properties of a 
single eigenstate
$\psi_\alpha ({\bf{r}})$\ , e.g., about the distribution 
$P(u)$ of the quantity $u\equiv |\psi_\alpha ({\bf{r}})|^2$.
Extensive studies of chaotic\cite{berry} and 
disordered\cite{disorder} cavities
have demonstrated that the main part of the distribution is 
described by the Porter-Thomas statistics
\begin{equation}
P_{\scriptscriptstyle \!\! P-T} (u) = 
\left(\frac{V}{2\pi u}\right)^{1/2} 
\exp (-uV/2)  \ \ ,
\end{equation}
where $V$ is the volume of the cavity and $\langle{u}\rangle =1/V$.

The above discussion suggests the following idea.
Let us assume that the cavity is weakly coupled to the 
environment, for instance, via a small opening or due to some small
absorption in the bulk. Then, by placing a monochromatic source
inside the cavity and tuning its frequency $\omega$ to a resonance,
one will generate in the cavity an intensity pattern $I({\bf{r}})$
which closely follows the "profile" $|\psi_\alpha ({\bf{r}})|^2$ of
the eigenstate $\alpha$  with frequency $\omega_\alpha\simeq\omega$.
Thus, the intensity distribution will be given by the 
Porter-Thomas statistics. On the other hand, for a sufficiently strong 
coupling (an open system) the distribution should obey the Rayleigh
statistics, Eq.~(1). The main purpose of this letter is to investigate 
the crossover between weak and strong coupling and to propose a 
generalized distribution $P(I)$ which interpolates between the
two regimes.

Let us first emphasize the difference between an open 
and a closed system, using the simple picture of addition of many random
waves\cite{ish}. In an open system the local field $\psi({\bf{r}},t)$ can be
viewed as a sum of great number of {\it traveling} waves, arriving to a 
point ${\bf{r}}$ from various scattering processes:
\begin{equation}
\psi({\bf{r}},t)= N^{-1/2}\sum_{n=1}^N
\cos(\theta_n + {\bf{k}}_n\cdot{\bf{r}} - \omega t ) \ \ ,
\end{equation}
where the phases $\theta_n$ are completely random and all the 
amplitudes have been taken to be equal (one could assume random
independent amplitudes, without any change in the results). The wave
vectors ${\bf{k}}_n$ are uniformly distributed on a $d$-dimensional
sphere $(d=2,3)$ of radius $k_o$. The instantaneous local intensity
is defined as $\psi^2({\bf{r}},t)$. The measured quantity, $I$, is the
intensity averaged over time, i.e. over one period $T=2\pi/\omega$\ :
\begin{equation}
I=\frac{1}{T}\int_0^T \!\! dt\,  \psi^2({\bf{r}},t) =
\frac{1}{2N}\sum_{n,m}^N \cos[ \theta_n - \theta_m + 
({\bf{k}}_n -{\bf{k}}_m)\cdot{\bf{r}}] \ \ .
\end{equation}
Note that the same expression for $I$ is obtained if one assumes a complex,
time independent field
\begin{equation}
\psi ({\bf{r}}) = (2N)^{-1/2}\sum_{n=1}^N 
\exp [i(\theta_n + {\bf{k}}_n\cdot{\bf{r}})]
\end{equation}
and defines the intensity as $I({\bf{r}})\equiv |\psi ({\bf{r}})|^2$.
It follows now from the central limit theorem that both $\Re \psi$
and $\Im \psi$ are independent Gaussian variables, with zero mean and
equal variances, which leads to Eq.~(1) for the intensity distribution.

In a closed system the field is viewed as a sum of many
{\it standing} waves:
\begin{equation}
\psi ({\bf{r}},t) = N^{-1/2}\sum_{n=1}^N 
\cos (\theta_n + {\bf{k}}_n\cdot{\bf{r}}) \cos\omega t
\end{equation}
As far as the time-averaged intensity is concerned, one can ignore
the factor \ $\cos\omega t$\ and define a stationary field
\begin{equation}
\psi ({\bf{r}}) = (2N)^{-1/2}\sum_{n=1}^N 
\cos (\theta_n + {\bf{k}}_n\cdot{\bf{r}}) \ \ .
\end{equation}
The central limit theorem now tells us that $\psi_\omega({\bf{r}})$ is a
Gaussian variable with zero mean, and the Porter-Thomas statistics
\begin{equation}
P_{\scriptscriptstyle \!\! P-T} (I) = \left(\frac{1}{2\pi\langle{I}\rangle I }\right)^{1/2} 
\exp (-I/2\langle{I}\rangle )
\end{equation}
for the intensity $I\equiv\psi_\omega^2 ({\bf{r}})$
follows immediately.

After clarifying the difference between sums of random
traveling waves (open systems) and standing waves (closed systems), we
move to the general case of a disordered cavity coupled to the external
world. We write the stationary field as
\begin{equation}
\psi ({\bf{r}}) = [2N(1+2\epsilon^2)]^{-1/2}\sum_{n=1}^N 
\left\{\cos (\theta_n + {\bf{k}}_n\cdot{\bf{r}}) +
\epsilon \exp [i(\theta^{\prime}_n + 
{\bf{k}}_n\cdot{\bf{r}})] \right\}
\end{equation}
where the parameter
$\epsilon$ describes the strength of the coupling.  
For small $\epsilon$, 
and when the source frequency is tuned to a resonance, the 
field consists of a large-amplitude standing wave (an eigenstate)
with a small traveling wave "riding" on top of it. The intensity
distribution
$P(I)$ is close to the expression  in Eq.~(9). Large $\epsilon$ 
corresponds to an open system, where the field is mostly a traveling
wave, and $P(I)$ is close to the Rayleigh distribution, Eq.~(1).

All the phases, $\theta_n$ and $\theta^{\prime}_n$, in Eq.~(10) 
are independent and uniformly distributed between $0$ and $2\pi$.
It is then clear that both $\Re \psi$ and 
$\Im \psi$ are independent Gaussian variables
with variances 
$\langle{(\Re\psi)^2}\rangle /\langle{(\Im\psi)^2}\rangle 
=(1+\epsilon^2)/\epsilon^2$.
This leads to the following distribution for the intensity
$I=(\Re\psi)^2+(\Im\psi)^2$:
\begin{equation}
P (I)=\frac{1 + 2\epsilon^2}
{2\langle{I}\rangle\epsilon\sqrt{1+\epsilon^2}}
\exp \left[-\frac{I}{4\langle{I}\rangle\epsilon^2}
\frac{ (1 + 2\epsilon^2 )^2}{1 +\epsilon^2}\right]
I_o \left[
\frac{I}{4\langle{I}\rangle\epsilon^2}\frac{1 + 2\epsilon^2}{1 +\epsilon^2}
\right]
\end{equation}
where $I_o[x]$ is the modified Bessel function. The n'th moment of this
distribution is then given by 
\mbox{$\langle{I^n}\rangle = \langle{I}\rangle^n\, n!\, _2F_1 \left[
{\scriptstyle -\frac{n}{2}, \frac{1-n}{2},1 , (1 + 2\epsilon^2)^{-2}}
\right]$}, where $_2F_1$ is the
Gaussian hypergeometric function.

Since the parameter \ $\epsilon$\ in Eq.~(10) is attached to
the propagating part of the field, it can be related to the
(averaged over time) current density\cite{frisch}:
\begin{equation}
{\bf{J}} ({\bf{r}})=\frac{ic}{2k_o} 
\left[
\psi ({\bf{r}})\nabla\psi^{\ast} ({\bf{r}}) - {\rm c.c}\ \right] \ \ .
\end{equation}
Substituting \ $\psi$\ from Eq.~(10) and averaging over phases, one 
finds that 
\mbox{$\langle{{\bf{J}}({\bf{r}})}\rangle$} vanishes\cite{foot}
and 
$\langle{J^2({\bf{r}})}\rangle =
2c^2\langle{I}\rangle^2\epsilon^2 (1 +\epsilon^2)/(1 +2\epsilon^2)^2$.
Therefore, instead of using the somewhat vague notion of the "coupling
strength $\epsilon$" for parametrization of the distribution $P(I)$,
one can use the dimensionless ratio
\mbox{$\delta\equiv \langle{J^2}\rangle /c^2\langle{I}\rangle^2$}.

We, thus, propose a one-parameter distribution $P_\delta (I)$ for the
intensity (we normalize $I$ to its average value,
i.e.\ choose $\langle{I}\rangle =1$):
\begin{equation}
P_\delta (I)= \frac{1}{\sqrt{2\delta}} \exp(-I/2\delta) 
I_o\left(I\sqrt{1-2\delta}/2\delta\right) \ \ .
\end{equation}
The parameter \ $\delta$\ can assume values from $0$ 
(closed system, no current)
to $1/2$ (open system, maximal current density). 
When this parameter changes
from $0$ to $1/2$, the intensity distribution changes 
from Porter-Thomas
to Rayleigh.

Let us mention that a somewhat different crossover phenomenon
has beenconsidered in Ref.\cite{cross}. These authors
discussed the statistics of 
$|\psi_\alpha({\bf{r}})|^2\equiv u$ for an electron's
eigenstate in a quantum dot, in the presence 
of an arbitrary magnetic field. For zero field the 
distribution $P(u)$ is given by Eq.~(3), whereas for 
sufficiently strong 
field it crosses over to a Rayleigh distribution
\mbox{$P(u)=V\exp(-Vu)$}. In this crossover problem,
as opposed to the one considered in the present paper, 
the system always remains closed.

In a similar way one can consider the distribution for the
local current density ${\bf{J}} ({\bf{r}})$ or the joint
probability distribution $P(I,{\bf{J}})$.
We will not discuss here these objects but limit the 
discussion to the joint probability distribution
$P(I_1, I_2)$, where $I_i\equiv I ({\bf{r}}_i ) \ \ (i=1,2)$. 
We start with a wave propagating in an open system.
The field $\psi ({\bf{r}})$ is then given by Eq.~(6).
It follows from
that equation that, in the
Large $N$ limit, the joint probability distribution for
$\psi ({\bf{r}}_1)\equiv\psi_1$ and
$\psi ({\bf{r}}_2)\equiv\psi_2$ is:
\begin{equation}
W(\psi_1,\psi_2) = \frac{1}{\pi^2 {\rm det} K}
\exp[-\psi^{\ast}_i (K^{-1})_{ij}\psi_j]
\end{equation}
where $K_{ij} = \langle{\psi_i\psi^{\ast}_j}\rangle$
is the $2\times 2$ covariance matrix with
$K_{11}=K_{22}=1$,$K_{12}=K^{\ast}_{21}= f(\rho)$,
and $\rho\equiv |{\bf{r}}_1 - {\bf{r}}_2 |$.
The explicit form of the field-field correlation function 
$f(\rho)$ will be given below. Transforming
to polar coordinates, \mbox{$\psi_i=\sqrt{I_i}\exp(i\phi_i )$},
and integrating out the phases, one obtains:
\begin{equation}
P(I_1,I_2)= \frac{1}{1- |f|^2} 
\exp\left(-\frac{I_1 +I_2}{1- |f|^2}\right)
I_o\left(\frac{2|f|\sqrt{I_1 I_2}}{1- |f|^2}\right)
\end{equation}
Eq.~(14) is the standard assumption in the theory of optical
and acoustical speckles and the resulting distribution
$P(I_1,I_2)$ is well known in optics and acoustics of
disordered media\cite{ish,ebe}.

After averaging the product
$\psi ({\bf{r}}_1)\psi^{\ast} ({\bf{r}}_2)$ over
the random phases $\theta_n$, one finds
$f(\rho)=\sum_n \exp (i {\bf{k}}_n\cdot{\bf{\rho}} )$
where the sum is taken over $N$ points on a unit sphere.
In the $N\rightarrow\infty$ limit, replacing the sum by
an integral, one finds $f(\rho)=J_o (k_o\rho)$ in 2 dimensions
and $f(\rho)=(k_o\rho)^{-1}\sin(k_o\rho)$
in 3 dimensions. (A more rigorous calculation\cite{bor}
shows that, for an open geometry, $f(\rho)$ decays
exponentially for $\rho$ larger than the mean free path $\ell$).

Eqs.~(14),(15) describe the statistics of radiation in
an open system.
In contrast, for a weakly coupled cavity (under resonance
condition) the main part of the field corresponds to
a standing wave. Such a field is represented by a sum of
real waves, Eq.~(8), and a derivation analogous to the
outlined above gives
\begin{equation}
P(I_1,I_2)= \frac{1}
{2\pi\sqrt{1-f^2}}\frac{1}{\sqrt{I_1 I_2}}
\exp\left(-\frac{I_1 +I_2}{2(1- f^2)}\right) 
\cosh \left(\frac{f\sqrt{I_1 I_2}}{1- f^2}\right)
\end{equation}
The statistics of eigenstates
$\psi_\alpha ({\bf{r}})$ in a closed system has
been rigorously studied by Prigodin and co-workers
with the help of a zero-dimensional supersymmetric non-linear
$\sigma$-model\cite{prigod1,prigod2}.
They studied the joint probability distribution $P(u_1,u_2)$, where
$u_i\equiv |\psi_\alpha ({\bf{r}}_i )|^2 \ \ (i=1,2)$.
For the unitary case\cite{prigod1} (broken time-reversal
symmetry), an expression identical to Eq.~(15) 
(with $I_i$ replaced by $u_i$) was obtained.
For the orthogonal case, Prigodin {\it et al.}\cite{prigod2}
ended up with a rather complicated expression, containing
a double integral. Later it has been shown by Srednicki\cite{mark}
that the expression in Ref.\cite{prigod2}
can be reduced to the function given in Eq.~(16). He was using the
assumption\cite{berry} that a chaotic wavefunction,
$\psi_\alpha ({\bf{r}})$,
obeys the statistics of a Gaussian random process.
This is in complete analogy with the standard assumption
of the speckle theory\cite{ish,ebe}, as outlined above.
Again, the difference is that in the speckle theory one
usually considers propagating waves in an open geometry,
whereas Refs.[10-12] study a single eigenstate in an isolated 
system.

Now, we can analyze the general case of a disordered
cavity coupled with arbitrary strength to the external world.
The local field is now given by a combination of 
traveling and standing waves, Eq.~(10). As a result,
the real and imaginary parts of the field at two points,
${\bf{r}}_1$ and ${\bf{r}}_2$,
are components of a four-dimensional Gaussian vector,
$\Phi^T=(\Re\psi_1, \Im\psi_1, \Re\psi_2, \Im\psi_2)$,
with the following covariance matrix:
\begin{equation}
K_{ij}\equiv\langle{\Phi_i\Phi_j}\rangle = 
\frac{1}{1+2\epsilon^2} \left[ \begin{array}{cccc}
1+\epsilon^2 & 0          & (1+\epsilon^2)f & 0 \\
0            & \epsilon^2 &   0            &\epsilon^2f \\
(1+\epsilon^2)f & 0     & 1+\epsilon^2     & 0 \\
0            &  \epsilon^2f & 0  & \epsilon^2
\end{array} \right]\rule{1.5cm}{0cm}
\end{equation}
After some lengthy algebra, this leads to 
\begin{eqnarray}
P_\delta (I_1,I_2)= \frac{\exp[-(I_1 + I_2)/2\delta]}{2\delta (1 -f^2)} 
\int^{2\pi}_0\!\! \frac{d\theta_1 d\theta_2}{(2\pi)^2}
\exp\left\{\frac{\sqrt{1-2\delta}}{2\delta (1 -f^2)}
\ \ \times \right. \rule{1.5cm}{0cm}  \\  \left.
\rule{1.5cm}{0cm}\left[ I_1\cos 2\theta_1 +
I_2\cos 2\theta_2 + 2f\sqrt{I_1 I_2}
\left( \cos (\theta_1 -\theta_2) - 
\frac{\cos (\theta_1 + \theta_2)}{\sqrt{1-2\delta}}\right)
\right] \right\}  \nonumber
\end{eqnarray}
Eq.~(18) interpolates
between a weakly coupled cavity (at the resonance) and an open system.

In conclusion, we consider statistics of radiation in a
disordered cavity coupled to the environment. The coupling can occur
via  an opening in the wall of the cavity or via absorption in
the bulk. For weak coupling, and when the source 
frequency $\omega$ is close to
an eigenfrequency $\omega_\alpha$, the wave generated
in the cavity is close to a (standing) eigenmode
$\psi_\alpha ({\bf{r}})$, with only a small admixture
of a traveling wave. The intensity statistics is defined by
the statistics of the eigenfunction  $\psi_\alpha ({\bf{r}})$.
For strong coupling the system becomesopen and we recover the
old results of the speckle theory for propagating waves.

There are many similarities between intensity correlations
in open random systems and correlations  in a single eigenstate
of a disordered cavity. There are also some differences.
It has been already mentioned that, in open systems, the
field-field correlations decays exponentially for distances
$\rho$ larger than the mean free path $\ell$. Therefore
correlations described by Eq.~(15) are of a short range character,
$\rho{\stackrel{<}\sim}\ell$. For
distances $\rho\gg\ell$ a rather different type of correlation,
due to diffusion, takes over\cite{long}.

Finally, let us mention that intensity distributions discussed above,
such as in Eqs.~(1), (9) or (13), apply only to the "bulk" of the
distributions. Tails of the distributions, corresponding to
very large or very small values of $I$,
will show significant deviations from the above given expressions
and will not be universal. Indeed, it is well known that,
both in open and closed systems, 
distributions for various quantities (conductance, density of states,
$\scriptstyle |\psi_\alpha ({\bf{r}})|^2$\ ) develop
log-normal tails\cite{tails}. This must
also be true for the intensity distribution $P(I)$ discussed in this paper.
For instance, for a point source placed at ${\bf{r}}=0$, the field 
$\psi_\omega ({\bf{r}})$ is just the Green's function
$G_\omega (0,{\bf{r}})$ and the intensity is
$I= | G_\omega (0,{\bf{r}}) |^2$. The Green's function can
be expanded in terms of the eigenfunctions $\psi_\alpha ({\bf{r}})$,
and the log-normal tail of $P(|\psi_\alpha |^2)$ are responsible 
for such tails in the intensity distribution.


This work was initiated in summer 1995, during the workshops organized
by the E.~Schr{\"o}dinger Institute in Vienna. We are grateful to the
organizers for their hospitality and to the participants,
B.~Altshuler and B.~Simons, for illuminating discussions and explanations
concerning this work. We are also indebted to A.~Genack, D.~Khmelnitskii,
E.~Kogan and I.~Lerner for useful discussions and correspondence.
The research was supported by the Fund for the promotion of research
at the Technion and by the Israel Science Foundation administered by
the Israel Academy of Sciences and Humanities.

\begin{small}

\end{small}
\end{document}